\documentclass[preprintnumbers,nofootinbib,preprint,12pt,showpacs]{revtex4}

\usepackage{amssymb,epsf}
\usepackage{graphicx}
\usepackage{latexsym}

\newcommand{\hM}{{\hat M}}
\newcommand{\hJ}{{\hat J}}
\newcommand{\vphi}{\varphi}

\begin{document}

{~}

\vspace{1cm}

\title{Extremal Charged Rotating Dilaton Black Holes\\ in Odd Dimensions}

 \vspace{1.5truecm}

\author{
{\bf Masoud Allahverdizadeh, Jutta Kunz}
}
\affiliation{
{
Institut f\"ur Physik, Universit\"at Oldenburg, Postfach 2503\\
D-26111 Oldenburg, Germany}
}
\author{
{\bf Francisco Navarro-L\'erida}
}
\affiliation{
{Departamento de F\'{\i}sica At\'omica, Molecular y Nuclear, Ciencias F\'{\i}sicas\\
Universidad Complutense de Madrid, E-28040 Madrid, Spain}
}

\begin{abstract}

Employing higher order perturbation theory,
we find a new class of charged rotating black hole solutions 
of Einstein-Maxwell-dilaton theory 
with general dilaton coupling constant. 
Starting from the Myers-Perry solutions,
we use the electric charge as the perturbative parameter,
and focus on extremal black holes
with equal-magnitude angular momenta in odd dimensions.
We perform the perturbations up to 4th order for black holes 
in 5 dimensions and up to 3rd order in higher odd dimensions.
We calculate the physical properties of these black holes
and study their dependence on the charge 
and the dilaton coupling constant.
\end{abstract}

\pacs{04.20.Ha, 04.20.Jb, 04.40.Nr, 04.50.-h, 04.70.Bw.}

\date{\today}

\maketitle 

\newpage

\section{Introduction}

Starting with the pioneering work of Kaluza and Klein,
the quest to unify the fundamental forces of nature has
led to higher dimensional theories,  
yielding additional fields upon reduction to lower dimensions.
In the low energy limit of the string theory, one recovers 
for instance, Einstein gravity along with a scalar dilaton field 
which is non-minimally coupled to matter fields
\cite{Green:1987sp}.

Exact solutions for charged dilaton black holes
where the dilaton is coupled to the Maxwell field 
have been considered by many authors.
This showed that the presence of the dilaton has important consequences 
for the black hole properties.
Gibbons and Maeda, for instance, obtained the general
family of static charged dilaton black holes in $D$ dimensions 
\cite{Gibbons:1987ps,Garfinkle:1990qj,Gregory:1992kr}.
They demonstrated that the physical properties of the
black holes depend sensitively on the value of the
dilaton coupling constant, with critical values of this
coupling constant separating qualitatively different
thermodynamical behaviour.

Exact rotating dilaton black hole solutions
have been obtained only for special values of the dilaton
coupling constant.
In 4 dimensions,
rotating charged dilaton black holes are known for
the Kaluza-Klein value of the coupling constant,
where solution generating techniques can be applied
\cite{Frolov:1987rj,Rasheed:1995zv,Larsen:1999pp}.
For general dilaton coupling constant,
only perturbative results for small angular momentum
\cite{Horne:1992zy,Shiraishi:1992jt}
or small charge \cite{Casadio:1996sj}
are available
\cite{Horne:1992zy,Casadio:1996sj}
as well as numerical results 
\cite{Kleihaus:2003df}.
Interestingly, already a small amount of rotation
leads to a qualitative change of the properties of
extremal black holes.
Moreover, the presence of both electric and magnetic charge
allows for rotating black holes with static \cite{Rasheed:1995zv}
and counterrotating horizons \cite{Kleihaus:2003df}.

In higher dimensions, the charged dilatonic generalizations
of the Myers-Perry black holes \cite{Myers:1986un}
are known for the respective Kaluza-Klein
value of the dilaton coupling constant, which depends
on the number of dimensions $D$.
Their domain of existence
and their properties were discussed in \cite{Kunz:2006jd}.
The inclusion of additional fields, as required by supersymmetry
or string theory, led to further exact solutions of higher
dimensional rotating dilatonic black holes 
\cite{Youm:1997hw,Llatas:1996gh,Horowitz:1995tm}.

Generalizing the lowest order perturbative Einstein-Maxwell results
for higher-dimensional charged rotating black holes
\cite{Aliev:2005npa,Aliev:2006yk}
to the case of Einstein-Maxwell-dilaton theory with
general dilaton coupling constant,
the properties of charged rotating dilaton black holes with
infinitesimally small angular momentum were investigated
in \cite{Sheykhi:2008bs}.
Also charged rotating dilaton black rings
were constructed perturbatively \cite{Kunduri:2004da}.
Rotating dilatonic black holes 
with the charge as the perturbative parameter have not yet been
obtained for general dilaton coupling.

In general, rotating black holes in $D$ dimensions possess
$[(D-1)/2]$ independent angular momenta \cite{Myers:1986un}.
When specializing to black holes with equal-magnitude angular momenta 
in odd dimensions, however,
the symmetry of the solutions enhances greatly.
This reduces the problem from cohomogeneity-$[(D-1)/2]$ to cohomogeneity-1,
and thus makes the solutions amenable to higher order
perturbative techniques. 

Recently, employing higher order perturbation theory
with the charge as the perturbative parameter,
the properties of such cohomogeneity-1 
rotating Einstein-Maxwell black holes were investigated
in five dimensions \cite{NavarroLerida:2007ez}.
Subsequently, this perturbative method was 
extended to obtain Einstein-Maxwell black holes 
with equal-magnitude angular momenta
in general odd dimensions,
focussing on extremal black holes
\cite{Allahverdizadeh:2010xx}.

These higher order perturbative results 
confirmed previous numerical studies
\cite{Kunz:2005nm,Kunz:2006eh},
showing that the lowest order perturbative results 
for the gyromagnetic ratio, $g=D-2$
\cite{Aliev:2004ec,Aliev:2005npa,Aliev:2006yk}
receive corrections, depending
upon the charge to mass ratio $Q/M$.
Indeed, in 3rd order the gyromagnetic ratio becomes
$g = (D-2) \left[1+\frac{1}{16} \left(\frac{Q}{M}\right)^2\right]$
\cite{Allahverdizadeh:2010xx}.
The higher order perturbative method was also employed to study
cohomogeneity-1 black holes in Einstein-Maxwell-Chern-Simons theory
\cite{Allahverdizadeh:2010xx},
allowing for further insight into these intriguing black holes
\cite{Breckenridge:1996sn,Breckenridge:1996is,Cvetic:2004hs,Chong:2005hr,Kunz:2005ei,Kunz:2006yp,Aliev:2008bh}.

The symmetry enhancement for black holes in odd dimensions with
equal-magnitude angular momenta 
is retained in the presence of a dilaton field.
In this paper we are thus led to apply the higher order perturbative method
to find charged rotating black holes in dilaton gravity.
Again, we focus on black holes at extremality.
Starting from the Myers-Perry black holes,
we evaluate the perturbative series
up to 4th order in the charge for black holes in 5 dimensions,
and up to 3rd order for black holes in more than 5 dimensions.
We determine the physical properties of these black holes
for general dilaton coupling constant. 
In particular, we investigate the effects of 
the presence of the dilaton field and
the perturbative parameter on the gyromagnetic ratio
of these rotating black holes.

The remainder of this paper is outlined as follows. 
In the next section, we present the metric,
the dilaton field and the gauge potential
for black holes in odd dimensions with 
equal-magnitude angular momenta. 
We introduce the perturbation series 
and present the global and horizon properties 
for these black holes. 
In section III we give the perturbative solution for extremal black holes 
of Einstein-Maxwell-dilaton theory in 5 dimensions. 
In section IV we present the generalization to general odd dimensions. 
We demonstrate the effect of the presence of the dilaton field and the charge
on the mass, the angular momentum, the magnetic moment
and the gyromagnetic ratio of these extremal rotating black holes. 
We also evaluate their horizon properties. 
Our conclusions are drawn in section V. 
The formulae for the metric, the dilaton field and the gauge
potential in $D$ dimensions are given in the Appendix.

\section{Black hole properties}

\subsection{Einstein-Maxwell dilaton theory}

We consider Einstein-Maxwell dilaton theory in $D$
dimensions, possessing the action
\begin{eqnarray}
S &=&\int d^{D}x\sqrt{-g}\left(
R\text{
}-\frac{1}{2}\partial_{\mu}\Phi \partial^{\mu}\Phi-\frac{1}{4}e^{-2h \Phi}F_{\mu \nu }F^{\mu \nu }\right),  \label{act1}
\end{eqnarray}
where ${R}$ is the scalar curvature, $\Phi$ is the dilaton field,
$F_{\mu \nu }=\partial _{\mu }A_{\nu }-\partial _{\nu }A_{\mu }$
is the electromagnetic field tensor, and $A_{\mu }$ is the
electromagnetic potential. $h$ is an arbitrary constant
governing the strength of the coupling between the dilaton and the
Maxwell field.
The units are chosen such that $16 \pi G_D=1$.

The equations of motion can be obtained by varying the action 
with respect to the metric $g_{\mu \nu}$, the dilaton field $\Phi $  
and the gauge potential $A_{\mu }$, yielding
\begin{equation}
G_{\mu\nu}=\frac{1}{2}\left[\partial_{\mu}\Phi \partial_{\nu}\Phi
-\frac{1}{2}g_{\mu \nu}\partial_{\rho}\Phi \partial^{\rho}\Phi
+e^{-2h \Phi}\left(F_{\mu\rho} {F_\nu}^\rho 
 - \frac{1}{4} g_{\mu \nu} F_{\rho \sigma} F^{\rho \sigma}\right)\right]
\ , \label{FE1}
\end{equation}
\begin{equation}
\nabla ^{2}\Phi   =-\frac{h}{2}e^{-2h \Phi}F_{\mu\nu} F^{\mu \nu} \ ,
\label{FE2}
\end{equation}
\begin{equation}
\nabla_\mu \left(e^{-2h \Phi}F^{\mu \nu}\right)  = 0 \ . \
\label{FE3}
\end{equation}

\subsection{Black holes in odd dimensions}

In general,
stationary black holes in $D$ dimensions possess $N = [(D-1)/2]$
independent angular momenta $J_{i}$ associated with
$N$ orthogonal planes of rotation \cite{Myers:1986un}.
($[(D-1)/2]$ denotes the integer part of $(D-1)/2$,
corresponding to the rank of the rotation group $SO(D-1)$.)
The general black holes solutions then fall into two classes,
into even-$D$ and odd-$D$-solutions \cite{Myers:1986un}.
Here we focus on perturbative charged rotating black holes in odd dimensions.

When all $N$ angular momenta have equal-magnitude,
the symmetry of the Myers-Perry black holes is strongly enhanced.
In odd dimensions, the symmetry then increases from
$R \times U(1)^N$ to $R \times U(N)$,
thus changing the problem from cohomogeneity-$N$ to cohomogeneity-1.
The original system of partial differential equations
in $N$ variables then reduces to a system of ordinary differential
equations.
In the presence of an electromagnetic field,
this symmetry enhancement is retained \cite{Kunz:2006eh},
and likewise in the presence of a dilaton field.
Since the angular dependence factorizes, the rotating
cohomogeneity-1 black holes of Einstein-Maxwell-dilaton theory
are amenable to higher order perturbation theory, 
as developed recently
\cite{NavarroLerida:2007ez,Allahverdizadeh:2010xx}
for Einstein-Maxwell and Einstein-Maxwell-Chern-Simons theory.

To obtain such perturbative charged dilatonic generalizations of the
$D$-dimensional Myers-Perry solutions \cite{Myers:1986un}, 
we employ the following parametrization for the metric
\cite{Kunz:2006eh,Kunz:2006yp}
\begin{eqnarray}\label{metric1}
\phantom{a=a}\nonumber \\
ds^2 &=&g_{tt}dt^2+\frac{dr^2}{W} + r^2\left[\sum^{N-1}_{i=1}\left(\prod^{i-1}_{j=0}\cos^{2}\theta_{j}\right)d\theta^{2}_{i}+\sum^{N}_{k=1}\left(\prod^{k-1}_{l=0}\cos^{2}\theta_{l}\right)\sin^{2}\theta_{k}d\varphi^{2}_{k}\right]
\nonumber \\[10pt]
&+&V\left[\sum^{N}_{k=1}\left(\prod^{k-1}_{l=0}\cos^{2}\theta_{l}\right)\sin^{2}\theta_{k}\varepsilon_{k}d\varphi_{k}\right]^{2}-2B\sum^{N}_{k=1}\left(\prod^{k-1}_{l=0}\cos^{2}\theta_{l}\right)\sin^{2}\theta_{k}\varepsilon_{k}d\varphi_{k}dt \  ,
\phantom{a=a}\nonumber \\
\end{eqnarray}
where $\theta_{0}\equiv0$, $\theta_{i}\in[0,\pi/2]$ 
for $i=1,...,N-1$, $\theta_{N}\equiv \pi/2$, 
$\varphi_{k}\in[0,2\pi]$ for $k=1,...,N$, 
and $\varepsilon_{k}=\pm1$ denotes the sense of rotation 
in the $k$-th orthogonal plane of rotation.
An adequate parametrization for the gauge potential is given by
\begin{eqnarray}\label{A1}
A_{\mu}dx^{\mu} &=& a_{t} dt
+a_{\varphi}\sum^{N}_{k=1}\left(\prod^{k-1}_{l=0}\cos^{2}\theta_{l}\right)
 \sin^{2}\theta_{k}\varepsilon_{k}d\varphi_{k} \ , 
\end{eqnarray}
where the metric functions $g_{tt}$, $W$, $V$, $B$, 
the functions for the gauge potential $a_{t}$, $a_{\varphi}$, 
and the dilaton function $\Phi$ 
depend only on the radial coordinate $r$.

\subsection{Perturbation theory}

We now consider perturbations about the Myers-Perry solutions,
with the electric charge as the perturbative parameter.
In the presence of the dilaton field
we obtain the perturbation series in the general form
\begin{eqnarray}\label{gtt}
g_{tt} = -1+\frac{2\hat{M}}{r^{D-3}}+q^{2}g^{(2)}_{tt}+q^{4}g^{(4)}_{tt}
+O(q^{6}) \ ,
\end{eqnarray}
\begin{eqnarray}\label{W}
W =1-\frac{2\hat{M}}{r^{D-3}}+\frac{2\hat{J}^{2}}{\hat{M}r^{D-1}}
+q^{2}W^{(2)}+q^{4}W^{(4)}+O(q^{6}) \ ,
\end{eqnarray}
\begin{eqnarray}\label{N}
V = \frac{2\hat{J}^{2}}{\hat{M}r^{D-3}}+q^{2}V^{(2)}+q^{4}V^{(4)}
+O(q^{6}) \ ,
\end{eqnarray}
\begin{eqnarray}\label{B}
B = \frac{2\hat{J}}{r^{D-3}}+q^{2}B^{(2)}+q^{4}B^{(4)}
+O(q^{6}) \ ,
\end{eqnarray}
\begin{eqnarray}\label{Phi}
\Phi = q^{2}\Phi^{(2)}+q^{4}\Phi^{(4)}
+O(q^{6}) \ ,
\end{eqnarray}
\begin{eqnarray}\label{a0}
a_{t} = q a^{(1)}_{t}
+q^{3} a^{(3)}_{t}+O(q^{5}) \ ,
\end{eqnarray}
\begin{eqnarray}\label{avarphi}
a_{\varphi} = q a^{(1)}_{\varphi} 
+q^{3} a^{(3)}_{\varphi}+O(q^{5}) \ ,
\end{eqnarray}
where $q$ is the perturbative parameter associated with the
electric charge (see Eq.~(\ref{charge}) below).
Because of charge reversal symmetry 
the perturbations contain only even powers of $q$ in the metric
and the dilaton, and only odd powers of $q$ in the gauge potential.

Inserting the metric Eq.~(\ref{metric1}) and
the gauge potential Eq.~(\ref{A1})
with the perturbation expansion Eqs.~(\ref{gtt}-\ref{avarphi})
into the field equations Eqs.~(\ref{FE1}-\ref{FE3}),
we then solve these equations order by order.

\subsection{Physical Quantities }

The mass $M$, the equal-magnitude angular momenta $|J_i|=J$, 
the dilaton charge $\Sigma$,
the electric charge $Q$, and the magnetic moments $\mu_{\rm mag}$ 
can be read off the asymptotic behavior of the metric,
the dilaton function, and the gauge potential
\cite{Kunz:2006eh,Kunz:2006yp}
\begin{eqnarray}\label{quantities}
&&g_{tt}=-1+\frac{\tilde{M}}{r^{D-3}}+... \ ,  \quad 
B=\frac{\tilde{J}}{r^{D-3}}+... \ , \quad  
\Phi=\frac{\tilde{\Sigma}}{r^{D-3}}+... \ ,
\nonumber\\
&&a_{t}=\frac{\tilde{Q}}{r^{D-3}}+... \ ,  \quad   
a_{\varphi}=\frac{\tilde{\mu}_{\rm mag}}{r^{D-3}}+... \ ,  
\end{eqnarray}
where
\begin{eqnarray}\label{quantities1}
\tilde{M}&=&\frac{1}{(D-2)A(S^{D-2})}M \ ,  
\nonumber \\
\tilde{J}&=&\frac{1}{2A(S^{D-2})}J \ ,\nonumber \\ 
\tilde{\Sigma}&=&\frac{1}{(D-3)A(S^{D-2})}\Sigma  \ ,\nonumber \\
\tilde{Q}&=&\frac{1}{(D-3)A(S^{D-2})}Q \ ,   
\nonumber \\
\tilde{\mu}_{\rm mag}&=&\frac{1}{(D-3)A(S^{D-2})}\mu_{\rm mag} \ ,
\end{eqnarray}
and $A(S^{D-2})$ is the area of the unit $(D-2)$-sphere. 
The definition of the magnetic moment is
gauge-invariant and refers to an asymptotic rest frame
\cite{Aliev:2004ec,Allahverdizadeh:2010xx}.
The gyromagnetic ratio $g$ is given by
\begin{eqnarray}\label{g}
g = \frac{2M\mu_{\rm mag}}{QJ}  \ .
\end{eqnarray}

The event horizon is located at $r=r_{\rm H}$. 
The horizon angular velocities $|\Omega_i|=\Omega$ 
can be defined by imposing the Killing vector field
\begin{eqnarray}\label{chi}
\chi = \xi+\Omega\sum^{N}_{k=1}\epsilon_{k}\eta_{k} \ ,
\end{eqnarray}
to be null on and orthogonal to the horizon
($\chi=\partial_t$ and $\eta_{k}=\partial_{\varphi_k}$).
The horizon electrostatic potential $\Psi_{\rm el, H}$ 
of these black holes is given by
\begin{eqnarray}\label{PsiH}
\Psi_{\rm el, H} = \left. (a_{t}+\Omega a_{\varphi}) \right|_{r=r_{\rm H}} \ ,
\end{eqnarray}
and the surface gravity $\kappa_{\rm sg}$ is defined by
\begin{eqnarray}\label{sg1}
\kappa^{2}_{\rm sg} = \left.
-\frac{1}{2}(\nabla_{\mu}\chi_{\nu})(\nabla^{\mu}\chi^{\nu})\right|_{r=r_{\rm H}} \ .
\end{eqnarray}

The black holes further satisfy the Smarr-like mass formula 
\cite{Kleihaus:2002tc,Kleihaus:2003df,Kunz:2006jd}
\begin{equation}
M = 2 \frac{D-2}{D-3} {\kappa_{\rm sg} A_{\rm H}} 
+ \frac{D-2}{D-3} N \Omega J  
+  2\Psi_{\rm el, H} Q +\frac{\Sigma}{h} \ . \label{smarr-like}
\end{equation}
With the relation
\begin{equation}
\frac{\Sigma}{h} = -\Psi_{\rm el, H} Q  \ , \label{Sig}
\end{equation}
the Smarr mass formula can be given its usual form
\cite{Gauntlett:1998fz}
\begin{equation}
M = 2 \frac{D-2}{D-3} {\kappa_{\rm sg} A_{\rm H}} 
+ \frac{D-2}{D-3} N \Omega J  
+\Psi_{\rm el, H} Q \ . \label{smarr}
\end{equation}
Since the relations Eq.~(\ref{Sig}) and ~(\ref{smarr}) must also
hold for the perturbation solutions, 
this provides a good check of consistency for the perturbative scheme.

\section{Charged dilaton black holes in 5 dimensions}

Here we first present the perturbative solutions,
obtained in 4th order in the charge,
for extremal black holes with equal-magnitude angular momenta
in Einstein-Maxwell-dilaton theory
for general dilaton coupling constant $h$.

We now perform perturbation theory in the charge to 4th order.
We impose the extremality condition on the black holes 
and fix the angular momentum $J$ for all orders.
We further impose a regularity condition at the horizon. 
This choice then fixes all integration constants
which appear in the scheme.

Introducing the parameter $\nu$ for the extremal
Myers-Perry solutions by
\begin{equation}
\hM=2\nu^2, \quad\quad
\hJ=2\nu^3, 
\label{nu5}
\end{equation}
we obtain for the metric and the dilaton field
the perturbation expansion up to 5th order
in the perturbative parameter $q$
and for the gauge potential functions up to 4th order 
\begin{eqnarray}
g_{tt} &=& -1 + \frac{4\nu^2}{r^2} +\frac{r^2-4\nu^2}{12\nu^2 r^4} q^2\nonumber \\[8pt]
&&+\left[\frac{(2-3h^2)(r^2-2\nu^2)^2}{216\nu^8r^4}\ln\left(1-\frac{2\nu^2}{r^2}\right)+\frac{(11-18h^2)\nu^2r^4-16(2-3h^2)\nu^4r^2+16\nu^6}{576\nu^8r^6}\right]q^4 \nonumber \\[8pt]
&&+O(q^6) \ ,
    \nonumber \\[8pt]
W&=&1-\frac{4\nu^2}{r^2}+\frac{4 \nu^4}{r^4} -\frac{r^2-2\nu^2}{12\nu^2
    r^4}q^2+\Bigg{\{}\frac{(2-3h^2)(r^2-2\nu^2)^3}{288\nu^{10}r^4}\ln\left(1-\frac{2\nu^2}{r^2}\right)\nonumber \\[8pt]
&&+\frac{12(2-3h^2)r^8-(121-186h^2)\nu^2r^6+(181-204h^2)\nu^4r^4-8(8-15h^2)\nu^6r^2+144h^2\nu^8}{1728\nu^8r^8}\Bigg{\}}q^4\nonumber \\[8pt]
&&+ O(q^6) \ ,
    \nonumber \\[8pt]
V&=&\frac{4\nu^4}{r^2} -\frac{(r^2+2\nu^2)}{6 r^4} q^2-\Bigg{\{}\frac{(2-3h^2)(r^2-2\nu^2)(3r^6+8\nu^6)}{864\nu^{10}r^4}\ln\left(1-\frac{2\nu^2}{r^2}\right)\nonumber \\[8pt]
&&+\frac{4(2-3h^2)r^8-4(2-3h^2)\nu^2r^6-7\nu^4r^4+16(1-3h^2)\nu^6r^2-16\nu^8}{576\nu^8r^6}\Bigg{\}}q^4+ O(q^6) \ ,
    \nonumber \\[8pt]
B&=&\frac{4 \nu^3}{r^2} - \frac{\nu}{3 r^4}  q^2-\Bigg{\{}\frac{(2-3h^2)(r^2-2\nu^2)(3r^4-6\nu^2r^2+16\nu^4)}{1728\nu^{9}r^4}\ln\left(1-\frac{2\nu^2}{r^2}\right)\nonumber \\[8pt]
&&+\frac{(2-3h^2)r^6-3(2-3h^2)\nu^2r^4+3(4-8h^2)\nu^4r^2-8\nu^6}{288\nu^7r^6}\Bigg{\}}q^4 
+O(q^6) \ ,
\nonumber \\[8pt]
\Phi &=& \frac{-h}{4\nu^2r^2}q^2-\Bigg{\{}\frac{h(2-3h^2)(r^2-2\nu^2)}{144\nu^{8}r^2}\ln\left(1-\frac{2\nu^2}{r^2}\right)+\frac{((3-6h^2)r^2-\nu^2)h}{96\nu^6r^4}\Bigg{\}}q^4 + O(q^6) \ ,
\nonumber \\[8pt]
a_{t} &=& \frac{1}{r^2}q+\Bigg{\{}\frac{(2-3h^2)(r^2-2\nu^2)}{72\nu^{6}r^2}\ln\left(1-\frac{2\nu^2}{r^2}\right)-\frac{(2+6h^2)\nu^2-(2-3h^2)r^2}{36\nu^4r^4}\Bigg{\}}q^3 + O(q^5) \ , \ \ \ \ 
\nonumber \\[8pt]
a_\vphi&=&-\frac{\nu}{r^2}q -\Bigg{\{}\frac{(2-3h^2)(r^4-4\nu^4)}{144\nu^{7}r^2}\ln\left(1-\frac{2\nu^2}{r^2}\right)\nonumber \\[8pt]
&&+\frac{(2-3h^2)r^4+(1-6h^2)\nu^2r^2-(4+12h^2)\nu^4}{72\nu^5r^4}\Bigg{\}}q^3 
+ O(q^5) \ . 
\label{solution_EMd}
\end{eqnarray}
Clearly, for vanishing dilaton coupling constant,
the previous results for rotating Einstein-Maxwell black holes
are recovered \cite{NavarroLerida:2007ez}.

We note, that for generic values of
the dilaton coupling constant $h$ logarithms appear in the expansion.
These logarithms disappear for the special value
\begin{equation}
h= h_{\rm KK}=\sqrt{ \frac{2}{3} } \ . \label{h5_def}
\end{equation}
This value precisely coincides with that value of the
dilaton coupling, for which the exact general rotating
black hole solutions are known \cite{Kunz:2006jd}.

This situation is thus analogous to the case
of Einstein-Maxwell-Chern-Simons theory \cite{Allahverdizadeh:2010xx},
where the logarithms disappear,
when the Chern-Simons coupling assumes its
supergravity value, for which the exact general rotating
black hole solutions are known \cite{Chong:2005hr}.
In fact, in both these special cases, the logarithms disappear at all orders. 

From the above expansions the physical properties
of the rotating dilaton black hole solutions can be extracted.
To 4th order the global physical quantities are given by
\begin{eqnarray}
M &=& 24\pi^2 \nu^2 + \frac{\pi^2}{2\nu^2} q^2 
   + \frac{\pi^2(1-6h^2)}{288\nu^6} q^4 + O(q^6) \ , \nonumber \\[5pt]
J &=& 16\pi^2 \nu^3 \ (\mbox{for any order}) \ , \nonumber \\[5pt]
\Sigma &=& -\frac{\pi^2h }{\nu^2} q^2-
\frac{\pi^2h(1-6h^2)}{72 \nu^6} q^4 + O( q^6) \ , \nonumber \\[5pt]
Q &=& 4\pi^2 q \ , \nonumber \\[5pt]
\mu_{\rm {mag}} &=& 4\pi^2 \nu q-
\frac{\pi^2(1+3h^2)}{18 \nu^3} q^3 + O( q^5) \
, 
\label{dilaton_charge_EMd}
\end{eqnarray}

\noindent
and the gyromagnetic ratio $g$ is given by
\begin{equation}
g=
3 +\frac{(1-6h^2)}{48\nu^4}q^2 + O(q^4) \ . \label{g_factor_extremal_EMd}
\end{equation}

\begin{figure}[h!]
\epsfxsize=7cm
\centerline{\includegraphics[angle=0,width=92mm]{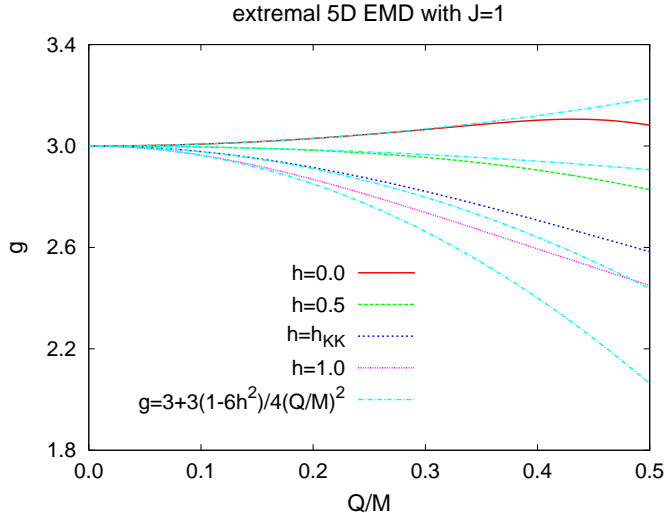}}
\caption{ 
The 3rd order perturbative values
for the gyromagnetic ratio $g$ versus the 
charge to mass ratio $Q/M$
for $D = 5$ Einstein-Maxwell-dilaton black holes
for several values of the dilaton coupling constant $h$ (cyan/dot-dashed).
For comparison, the analytically known result
($h_{\rm KK}$) and exact numerical results ($h=0$, 0.5, 1.0)
are also shown.
}
\label{fig1}
\end{figure}

Recently, it was shown that the presence of the dilaton field 
modifies the gyromagnetic ratio $g$,
by applying lowest order perturbation theory
with respect to the angular momentum
and thus evaluating $g$ in the slowly rotating case 
\cite{Sheykhi:2008bs}.
Here, we can see that the dilaton field generically modifies the
value of the gyromagnetic ratio in 2nd order 
in the charge, unless $h^2= h_{\rm cr}^2=1/6$.
For $h<h_{\rm cr}$ the gyromagnetic ratio grows with
increasing charge to mass ratio $Q/M$,
whereas for $h>h_{\rm cr}$ it decreases with increasing $Q/M$,
as seen in Fig.~\ref{fig1}.

We compare in Fig.~\ref{fig1} the 3rd order perturbative results
for the gyromagnetic ratio $g$ with the analytically known result
for the Kaluza-Klein case ($h_{\rm KK}$)
and numerical results for other values of the dilaton
coupling constant ($h=0$, 0.5, 1.0).
We note, that for small $h$ ($h=0$, 0.5)
the perturbative results are rather
good up to about one third of the domain of existence.
Note, that the range of possible values of the charge to mass ratio $Q/M$
for these 5-dimensional black holes
is bounded by $(Q/M)_{\rm max}$, where  
\begin{equation}
\left(\frac{Q}{M}\right)_{\rm max} = \sqrt{\frac{1}{3} + h^2} \ .
\end{equation}
For larger values of $h$ ($h_{\rm KK}$, 1.0)
the range of validity of the 3rd order perturbative results
decreases.

Let us now turn to the horizon properties of these black holes.
Their event horizon is located at $r=r_{\rm H}$, where
\begin{equation}
r_{\rm H} = \sqrt{2} \nu + \frac{\sqrt{2}}{96 \nu^3} q^2 
+\frac{\sqrt{2}(11-48h^2)}{18432 \nu^7} q^4
+ O( q^6) \ . \label{hor_rad_EMd}
\end{equation}
The horizon angular velocity $\Omega$,
the horizon area $A_{\rm H}$,
and the horizon electrostatic potential $\Psi_{\rm el, H}$
are given by
\begin{eqnarray}
&&\Omega = \frac{1}{2\nu} - \frac{1}{96 \nu^5} q^2 
-\frac{(1-6h^2)}{4608 \nu^9}q^4
+ O(q^6) \ , 
\nonumber \\
&&A_{\rm H} = 8 \pi^2 \nu^3 + O(q^6) \ ,
\nonumber \\
&&\Psi_{\rm el, H}=\frac{1}{4\nu^2} q 
+\frac{(1-6h^2)}{288\nu^6}q^3
+O(q^5) \ , \label{AH_and_PsiH_EMd}
\end{eqnarray}
and the surface gravity $\kappa_{\rm sg}$
vanishes for these extremal solutions.

\section{Charged black holes in odd dimensions}

To obtain the perturbative Einstein-Maxwell-dilaton
black holes in the general case of $D$ odd dimensions,
we proceed as in 5 dimensions. 
We fix the angular momentum for any perturbative order, 
and impose the extremality condition for all orders. 
We then introduce the parameter $\nu$ 
for the extremal Myers-Perry solutions in $D$ dimensions by 
\begin{equation}
\hat{M}=\frac{(D-1)^{\frac{(D-1)}{2}}\nu^{D-3}}{4(D-3)^{\frac{(D-3)}{2}}}\ ,
\quad\quad
\hat{J}=\frac{(D-1)^{\frac{(D-1)}{2}}\nu^{D-2}}{4(D-3)^{\frac{(D-3)}{2}}}\ .
\label{nu_gen}
\end{equation}
To fix all of the integration constants, we again need to make use of the
regularity condition of the horizon.
We employ the Smarr relation as a consistency check
of the new perturbative black hole solutions.

The perturbative expansion for the metric, the dilaton field,
and the gauge potential is exhibited in Appendix A
for generic values of the dilaton coupling constant $h$.
These expressions hold for the metric 
and the dilaton field in 3rd order,
and for the gauge potential in 4th order. 
We note, that for generic values of $h$
the higher order expansions 
contain non-rational functions.
As in 5 dimensions, however,
the expansions simplify considerably 
when $h$ assumes the special dimension-dependent value
\begin{equation}
h_{\rm KK}=\frac{D-1}{\sqrt{2(D-1)(D-2)}} \ , \label{h_def}
\end{equation}
for which the black hole solutions are known analytically \cite{Kunz:2006jd}.
Since the Einstein equations are modified by the presence of the dilaton
  in 4th order, the expansions for the metric functions given in the Appendix
  (namely, Eqs.~(\ref{ggtt}-\ref{BB})) do not depend on the dilaton
  coupling constant $h$. Concerning the dilaton field $\Phi$ the corresponding
  expression for the Kaluza-Klein solutions is nothing but Eq.~(\ref{PhiPhi})
  with $h$ restricted to Eq.~(\ref{h_def}). At the order of the  perturbations
  we 
  are treating in this paper, the main simplification for $h=h_{\rm KK}$
  occurs for the gauge potential functions. The complicated integral
  expressions appearing in Eqs.~(\ref{a00}-\ref{S3}) reduce to the following
  simple formulae for the Kaluza-Klein solutions:
\begin{eqnarray}
a_t &=& \frac{q}{r^{D-3}} - \frac{1}{D-2}
\left(\frac{D-3}{D-1}\right)^{\frac{D-3}{2}} \frac{1}{r^{2(D-3)}\nu^{D-3}} q^3
+ O(q^{5}) \ ,  \label{a0_KK} \\
a_\varphi &=& -\frac{\nu q}{r^{D-3}}  + 
\frac{(D-3)^{\frac{D-3}{2}}}{(D-2)(D-1)^{\frac{D-3}{2}}} \left[
\frac{1}{r^{2(D-3)}\nu^{D-4}} +
\frac{2(D-3)^{\frac{D-3}{2}}}{(D-1)^{\frac{D-1}{2}}} \frac{1}{r^{D-3}
  \nu^{2D-7}} \right] q^3 \nonumber \\
&& + O(q^5) \ . \label{aphi_KK}
\end{eqnarray}

\begin{figure}[h!]
\epsfxsize=7cm
\centerline{\includegraphics[angle=0,width=92mm]{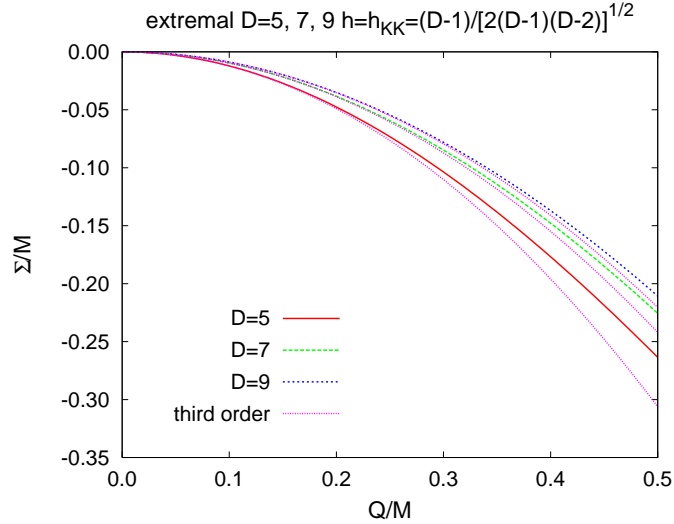}}
\caption{
The 3rd order perturbative values
for the dilaton charge $\Sigma$ versus the 
charge to mass ratio $Q/M$
for Einstein-Maxwell-dilaton black holes in 5, 7 and 9 dimensions
for the dilaton coupling constant $h_{\rm KK}$ (purple/dotted).
For comparison, the analytically known results are also shown.
}
\label{fig2}
\end{figure}

For a generic value of the dilaton coupling constant,
by extracting the asymptotic behavior of the solutions
Eqs.~(\ref{ggtt}-\ref{aphiphi})
from the perturbative expansion,
we now obtain the global properties for these Einstein-Maxwell-dilaton 
black holes
\begin{eqnarray}\label{charge}
M &=& A(S^{D-2})\left[\frac{\nu^{(D-3)}(D-2)(D-1)^{\frac{D-1}{2}}}{2(D-3)^{\frac{D-3}{2}}}+\frac{q^{2}(D-3)^{\frac{D-1}{2}}}{\nu^{D-3}(D-1)^{\frac{D-1}{2}}}\right]+O(q^{4}) \ ,
\nonumber \\[9pt]
J &=& \frac{A(S^{D-2})\nu^{D-2}(D-1)^{\frac{D-1}{2}}}{(D-3)^{\frac{D-3}{2}}} \ ,
\nonumber \\[9pt]
\Sigma &=& -\frac{2A(S^{D-2})(D-3)^{\frac{D-1}{2}}h }{\nu^{D-3}(D-1)^{\frac{D-1}{2}}} q^2+ O( q^4) \ , 
\nonumber \\[9pt]
Q &=& A(S^{D-2})(D-3)q \ ,
\nonumber \\[9pt]
\mu_{\rm mag} &=& A(S^{D-2})(D-3)\left[q\nu-\frac{q^3(D-3)^{D-3}(D-3+2(D-2)h^2)}{\nu^{2D-7}(D-2)^{2}(D-1)^{D-2}}\right]+O(q^{5}) \ .
\end{eqnarray}
The dilaton charge $\Sigma$ is exhibited in Fig.~\ref{fig2}
versus the 
charge to mass ratio $Q/M$
for Einstein-Maxwell-dilaton black holes in 5, 7 and 9 dimensions
for the dilaton coupling constant $h_{\rm KK}$,
where the solutions are known analytically \cite{Kunz:2006jd}.
Comparison shows, that the interval, where
the 3rd order perturbative results
are rather good, increases considerably with increasing dimension.
(For the dilaton coupling constant $h_{\rm KK}$ the range of possible values 
of the charge to mass ratio $Q/M$
is bounded by $Q/M\le 1$ in all dimensions.)

\begin{figure}[h!]
\epsfxsize=7cm
\centerline{\includegraphics[angle=0,width=92mm]{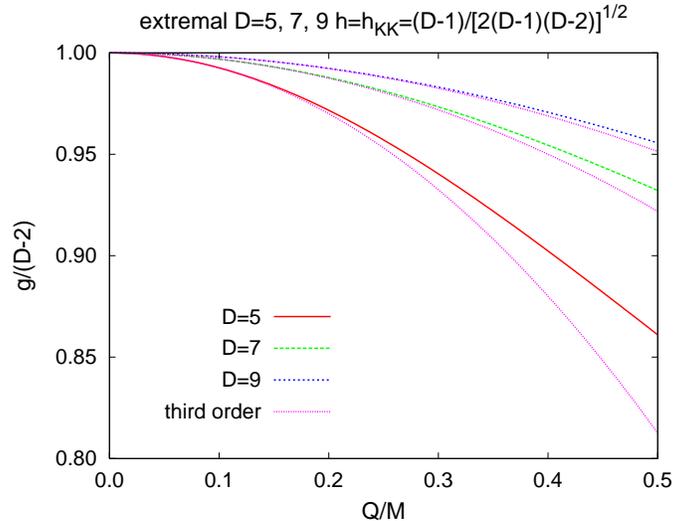}}
\caption{
The 3rd order perturbative values
for the gyromagnetic ratio $g$ versus the 
charge to mass ratio $Q/M$
for Einstein-Maxwell-dilaton black holes in 5, 7 and 9 dimensions
for the dilaton coupling constant $h_{\rm KK}$ (purple/dotted).
For comparison, the analytically known results are also shown.
}
\label{fig3}
\end{figure}

The gyromagnetic ratio $g$, in particular, is given by
\begin{eqnarray}\label{g2}
g = D-2+q^{2}\frac{(D-3)^{D-3}((D-3)^2-2(D-1)(D-2)h^2)}{(D-2)(D-1)^{D-1}\nu^{2(D-3)}}+O(q^{4}) \ .
\end{eqnarray}
Clearly, the gyromagnetic ratio of these
fast rotating black holes is modified
from its lowest order value $g=D-2$ 
in higher order perturbation theory, with appreciable
contributions coming from the dilaton field (proportional to $h^2$).
The gyromagnetic ratio is exhibited in Fig.~\ref{fig3}
versus the 
charge to mass ratio $Q/M$
for Einstein-Maxwell-dilaton black holes in 5, 7 and 9 dimensions
for the dilaton coupling constant $h_{\rm KK}$
and compared to the analytically known results \cite{Kunz:2006jd}.
As for the dilaton charge, we note that
the 3rd order perturbative results
are the better the higher the dimension.

The event horizon of these rotating dilaton black holes is located at
\begin{equation}
r_{\rm H}=\sqrt{\frac{D-1}{D-3}}\nu+\frac{(D-3)^{D-\frac{5}{2}}}{(D-2)(D-1)^{D-\frac{3}{2}}\nu^{2D-7}}q^2+O(q^{4}) \ . \label{r_H_odddim}
\end{equation}
The horizon angular velocity $\Omega$, and the horizon area $A_H$
are given by
\begin{eqnarray}\label{Omega}
&&\Omega = \frac{D-3}{\nu(D-1)}
-\frac{2q^2(D-3)^{D-1}}{\nu^{2D-5}(D-2)(D-1)^{D}}+O(q^{4}) \ ,
\nonumber \\
%
&&A_{\rm H}=\frac{\sqrt{2}A(S^{D-2})(D-1)^{\frac{(D-1)}{2}}\nu^{D-2}}{2(D-3)^{\frac{(D-2)}{2}}}+O(q^{4}) \ ,
\end{eqnarray}
and the surface gravity $\kappa_{\rm sg}$ vanishes
for these extremal solutions.


\section{Summary and Conclusion}

Focussing on odd dimensions, which allow for a cohomogeneity-1 
reduction, we have presented new 
perturbative solutions for extremal charged rotating
dilaton black holes with equal-magnitude angular momenta. 
These solutions are asymptotically flat
and their horizon has spherical topology.

Our strategy for obtaining these solutions
was based on the perturbative method,
where we solved the equations of motion
up to at least 3rd order in the perturbative parameter,
which we chose proportional to the charge.
In particular, we started from the
rotating black hole solutions in higher dimensions
\cite{Myers:1986un},
and considered the effect of adding a small amount
of charge to the solutions.
We then evaluated how the perturbative parameter and the dilaton field 
modify the physical properties of the solutions.

In 5 dimensions, we derived in this way
the metric and the dilaton field up to the 5th order 
in the perturbative parameter, and the gauge potential
up the 4th order.
We obtained the physical properties of 
these extremal charged rotating black holes
for general values of the dilaton coupling constant $h$. 
In particular,
we investigated the effect of the presence of the dilaton field
on the gyromagnetic ratio of these rapidly rotating black holes.

In $D>5$ odd dimensions, 
we presented the 3rd order perturbative expansion
for the metric and the dilaton field,
and the 4th order expansion for the gauge potential
for general values of the dilaton coupling constant,
and investigated their physical properties.
Comparison with the properties of the available analytical solutions
\cite{Kunz:2006jd} and with numerical results (in 5 dimensions)
demonstrated that
the range of validity of these 3rd order perturbative results
covers a considerable portion of the domain of existence
of these solutions.
Interestingly, our results reveal that the 
new perturbative solutions
are the better the higher the dimension.

The generalization to asymptotically nonflat dilaton 
black holes will represent the next important step.
While asymptotically (anti-)de Sitter vacuum black holes
are known \cite{Gibbons:2004js},
their generalizations to Einstein-Maxwell theory are known only
either numerically
\cite{Kunz:2007jq,Brihaye:2007vm,Brihaye:2008br}
or in the limit of slow rotation
\cite{Aliev:2006tt,Aliev:2007qi}.
Slowly rotating (anti-)de Sitter black holes
were also considered with a dilaton field included
\cite{Sheykhi:2008rm,Sheykhi:2009vb}.

In the presence of Liouville-type dilatonic potentials
perturbative black hole solutions were also obtained
for small angular momentum. 
These may be asymptotically anti-de Sitter \cite{Ghosh:2007jb}
or neither asymptotically flat nor (anti-)de Sitter
\cite{Ghosh:2002ut,Sheykhi:2006jh,Sheykhi:2006ee}.
Also rotating dilaton black rings with such
unusual asymptotics were obtained
\cite{Yazadjiev:2005aw}.
Moreover, dilaton black holes with squashed horizons
could be constructed \cite{Yazadjiev:2006iv,Allahverdizadeh:2009ay}.
Known only in the limit of slow rotation,
application of the present method will allow 
to obtain their fast rotating counterparts.

\acknowledgments{FNL gratefully acknowledges Minis\-terio de Ciencia e
  Innovaci\'on of Spain for financial support under project FIS2009-10614.}

\appendix

\section{} 

Here we give the perturbative expressions 
for the metric and the gauge potential in
Einstein-Maxwell-dilaton theory for general odd $D$.
The solutions up to 3rd order read
\begin{eqnarray}\label{ggtt}
g_{tt} &=& -1+\frac{(D-1)^{\frac{(D-1)}{2}}\nu^{D-3}}{2(D-3)^{\frac{(D-3)}{2}}r^{D-3}}\nonumber \\
&&-q^{2}\left[\frac{(D-3)}{2(D-2)r^{2(D-3)}}-\frac{(D-3)^{\frac{D-1}{2}}}{(D-2)(r\nu)^{D-3}(D-1)^{\frac{D-1}{2}}}\right]+O(q^{4}) \  ,
\end{eqnarray}
\begin{eqnarray}\label{WW}
W &=&1-\frac{(D-1)^{\frac{(D-1)}{2}}\nu^{D-3}}{2(D-3)^{\frac{(D-3)}{2}}r^{D-3}}+\frac{(D-1)^{\frac{(D-1)}{2}}\nu^{D-1}}{2(D-3)^{\frac{(D-3)}{2}}r^{D-1}} -\frac{q^{2}}{2(D-2)}\Bigg{\{}\frac{2(D-3)^{\frac{(D-1)}{2}}}{(r\nu)^{D-3}(D-1)^{\frac{(D-1)}{2}}}\nonumber \\
&&+\frac{(D-3)^{\frac{(D-3)}{2}}}{\nu^{D-5}r^{D-1}(D-1)^{\frac{(D-3)}{2}}}+\frac{(D-5)\nu^{2}}{r^{2(D-2)}}-\frac{D-3}{r^{2(D-3)}}\Bigg{\}}+O(q^{4}) \ ,
\end{eqnarray}
\begin{eqnarray}\label{VV}
V &=& \frac{(D-1)^{\frac{(D-1)}{2}}\nu^{D-1}}{2(D-3)^{\frac{(D-3)}{2}}r^{D-3}}\nonumber \\
&&-q^{2}\left[\frac{(D-3)\nu^2}{2(D-2)r^{2(D-3)}}+\frac{(D-3)^{\frac{D-3}{2}}}{(D-2)(D-1)^{\frac{D-3}{2}}\nu^{D-5}r^{D-3}}\right]+O(q^{4}) \ ,
\end{eqnarray}
\begin{eqnarray}\label{BB}
B = \frac{(D-1)^{\frac{(D-1)}{2}}\nu^{D-2}}{2(D-3)^{\frac{(D-3)}{2}}r^{D-3}}-\frac{\nu(D-3)q^{2}}{2(D-2)r^{2(D-3)}}+O(q^{4}) \ ,
\end{eqnarray}
\begin{eqnarray}\label{PhiPhi}
\Phi = \frac{-2h(D-3)^{\frac{(D-3)}{2}}}{\nu^{D-3}(D-1)^{\frac{(D-1)}{2}}r^{D-3}}q^{2}+O(q^{4}) \ ,
\end{eqnarray}
\begin{eqnarray}\label{a00}
a_{t} &=& \frac{q}{r^{D-3}}+q^3\int_{\infty}^{r}\frac{1}{\nu^{D-3}z^{3(D-2)}}\Bigg{\{}\frac{(D-3)^{D-1}z^{2(D-2)}}{(D-2)2^{\frac{D-7}{2}}(D-1)^{\frac{D-1}{2}}} \times I_{1}-S_{2}\times I_{4}\Bigg{\}}dz\nonumber \\
&&+O(q^{5}) \ ,
\end{eqnarray}
\begin{eqnarray}\label{aphiphi}
\hspace*{-2.5cm}
a_{\varphi}&=&\frac{-\nu q}{r^{D-3}}+q^{3}r^{2}\Bigg{\{}\int_{\infty}^{r}\frac{1}{\left[\nu^{2D-7}z^{4D-5}-\frac{\nu^{3D-10}(D-1)^{\frac{D-1}{2}}z^{3D-2}}{2(D-3)^{\frac{D-3}{2}}}+\frac{(D-1)^{\frac{D-1}{2}}\nu^{3D-8}z^{3D-4}}{2(D-3)^{\frac{D-3}{2}}}\right]}\nonumber \\
&&\Bigg{\{}\frac{\nu^{D-3}(D-3)^{D-2} z^{2(D-2)}}{2^{D-6}(D-2)(D-1)^{\frac{D-3}{2}}}\left(z^{D-1}-\frac{\nu^{D-3}(D-1)^{\frac{D-1}{2}}z^{2}}{2(D-3)^{\frac{D-3}{2}}}+\frac{\nu^{D-1}(D-1)^{\frac{D-1}{2}}}{2(D-3)^{\frac{D-3}{2}}}\right)\times I_{3} \nonumber \\
&&-2^{\frac{D-7}{2}}(D-1)\nu^{D-3}\left(\sum^{\frac{D-5}{2}}_{i=0}(i+1)\nu^{2i}(\frac{D-1}{2})^{i}(\frac{D-3}{2})^{\frac{D-2i-7}{2}}x^{D-2i-5}\right)
\nonumber \\
&& 
\times
\frac{((D-3)z^{2}-(D-1)\nu^{2})^{2}}{(D-3)^{\frac{D-1}{2}}}S_{2}\times I_{4} 
+S_{3}\Bigg{\}}dz\Bigg{\}}+O(q^{5}) \ ,
\end{eqnarray}
where in the above equations $I_{1}$, $I_{2}$, $I_{3}$, $I_{4}$, $S_{1}$, $S_{2}$ and $S_{3}$ are

\begin{eqnarray}\label{I1}
I_{1} &=& \int_{\infty}^{z}\left(\frac{1}{y^{D-2}}-\frac{\nu^{D-3}(D-1)^{\frac{D-1}{2}}}{2(D-3)^{\frac{D-3}{2}}y^{2D-5}}+\frac{(D-1)^{\frac{D-1}{2}}\nu^{D-1}}{2(D-3)^{\frac{D-3}{2}}y^{2D-3}}\right)\times S_{1}\times I_{2}dy \ ,
\end{eqnarray}

\begin{eqnarray}\label{S1}
S_{1} &=& \nu^{D-1}(\frac{D-1}{2})^{\frac{D+1}{2}}+h^2(D-2)(\frac{D-1}{2})^{\frac{D-1}{2}}\nu^{D-1}(D-3)\nonumber \\
&&-h^2\nu^{D-3}(D-2)(D-4)y^2-h^2(D-2)(\frac{D-3}{2})^{\frac{D-3}{2}}y^{D-1}\ ,
\end{eqnarray}
  
\begin{eqnarray}\label{I2}
I_{2} = \int_{\infty}^{y}\frac{x^{D-2}}{\left(\frac{(D-3)x^2}{2}-\frac{(D-1)\nu^{2}}{2}\right)^{4}\left(\sum^{\frac{D-5}{2}}_{i=0}(i+1)\nu^{2i}(\frac{D-1}{2})^{i}(\frac{D-3}{2})^{\frac{D-2i-7}{2}}x^{D-2i-5}\right)^{2}}dx \ ,
\end{eqnarray}

\begin{eqnarray}\label{S2}
S_{2} &=&\frac{1}{2^{D-3}}\bigg{\{}\frac{4h^{2}(D-3)^{\frac{3D-5}{2}}z^{2D-2}}{(D-1)^{\frac{D-1}{2}}}-\frac{\nu^{2}(D-3)^{\frac{3D-7}{2}}(\frac{D-1}{2(D-2)}-h^{2})z^{2D-4}}{(D-2)(D-1)^{\frac{D-5}{2}}}-4h^{2}\nu^{D-3}(D-3)^{D-1}z^{D+1}\nonumber \\
&&+4\nu^{D-1}(D-4)(D-3)^{D-2}(\frac{D-1}{2(D-2)(D-4)}+h^{2})z^{D-1}+h^{2}\nu^{2D-6}(D-1)^{\frac{D-1}{2}}(D-3)^{\frac{D+1}{2}}z^{4}\nonumber \\
&&-\nu^{2D-4}(2D-7)(D-1)^{\frac{D-1}{2}}(D-3)^{\frac{D-1}{2}}(\frac{D-1}{2(D-2)(2D-7)}+h^{2})z^{2}\nonumber \\
&&+\frac{\nu^{2D-2}(D-1)^{\frac{D-1}{2}}(D-3)^{\frac{D+3}{2}}(\frac{D-1}{2(D-2)(D-3)}+h^{2})}{D-2}\bigg{\}}  \ ,
\end{eqnarray}

\begin{eqnarray}\label{I3}
I_{3}&=&\int_{\infty}^{z}\frac{\left(y^{D-1}-\frac{\nu^{D-3}(D-1)^{\frac{D-1}{2}}y^{2}}{2(D-3)^{\frac{D-3}{2}}}+\frac{\nu^{D-1}(D-1)^{\frac{D-1}{2}}}{2(D-3)^{\frac{D-3}{2}}}\right)\times S_{1}\times I_{2}}{y^{2D-3}}dy
\ ,
\end{eqnarray}

\begin{eqnarray}\label{I4}
I_{4} = \int_{\infty}^{z}\frac{x^{D-2}}{\left(\frac{(D-3)x^2}{2}-\frac{(D-1)\nu^{2}}{2}\right)^{4}\left(\sum^{\frac{D-5}{2}}_{i=0}(i+1)\nu^{2i}(\frac{D-1}{2})^{i}(\frac{D-3}{2})^{\frac{D-2i-7}{2}}x^{D-2i-5}\right)^{2}}dx \ ,
\end{eqnarray}

\begin{eqnarray}\label{S3}
S_{3}&=&\frac{-2(D-3)^{D-3}(h^{2}+\frac{D-3}{2(D-2)})z^{3D-5}}{(D-2)(D-1)^{D-3}}+\frac{8\nu^{D-3}(D-3)^{\frac{D-3}{2}}(h^{2}+\frac{(D-1)(D-3)}{8(D-2)})z^{2D-2}}{(D-1)^{\frac{D-1}{2}}}\nonumber \\
&&+\frac{\nu^{D-1}(D-3)^{\frac{D-3}{2}}z^{2D-4}}{(D-2)(D-1)^{\frac{D-5}{2}}}-2h^{2}\nu^{2D-6}z^{D+1}+\frac{2\nu^{2D-4}(D-3)(h^{2}+\frac{D-1}{2(D-2)(D-3)})z^{D-1}}{(D-2)}  \ . 
\nonumber\\
\end{eqnarray}




\begin{thebibliography}{999}

\bibitem{Green:1987sp}
  M.~B.~Green, J.~H.~Schwarz and E.~Witten,
{\it  Cambridge, UK: Univ. Pr. (1987) 
(Cambridge Monographs On Mathematical Physics)}

\bibitem{Gibbons:1987ps}
  G.~W.~Gibbons and K.~i.~Maeda,
  Nucl.\ Phys.\  B {\bf 298}, 741 (1988).

\bibitem{Garfinkle:1990qj}
  D.~Garfinkle, G.~T.~Horowitz and A.~Strominger,
  Phys.\ Rev.\  D {\bf 43}, 3140 (1991)
  [Erratum-ibid.\  D {\bf 45}, 3888 (1992)].

\bibitem{Gregory:1992kr}
  R.~Gregory and J.~A.~Harvey,
  Phys.\ Rev.\  D {\bf 47}, 2411 (1993)
  [arXiv:hep-th/9209070].

\bibitem{Frolov:1987rj}
  V.~P.~Frolov, A.~I.~Zelnikov and U.~Bleyer,
  Annalen Phys.\  {\bf 44}, 371 (1987).

\bibitem{Rasheed:1995zv}
  D.~Rasheed,
  Nucl.\ Phys.\  B {\bf 454}, 379 (1995)
  [arXiv:hep-th/9505038].

\bibitem{Larsen:1999pp}
  F.~Larsen,
  Nucl.\ Phys.\  B {\bf 575}, 211 (2000)
  [arXiv:hep-th/9909102].

\bibitem{Horne:1992zy}
  J.~H.~Horne and G.~T.~Horowitz,
  Phys.\ Rev.\  D {\bf 46}, 1340 (1992)
  [arXiv:hep-th/9203083].

\bibitem{Shiraishi:1992jt}
  K.~Shiraishi,
  Phys.\ Lett.\  A {\bf 166}, 298 (1992).

\bibitem{Casadio:1996sj}
  R.~Casadio, B.~Harms, Y.~Leblanc and P.~H.~Cox,
  Phys.\ Rev.\  D {\bf 55}, 814 (1997)
  [arXiv:hep-th/9606069].

\bibitem{Kleihaus:2003df}
  B.~Kleihaus, J.~Kunz and F.~Navarro-Lerida,
  Phys.\ Rev.\  D {\bf 69}, 081501 (2004)
  [arXiv:gr-qc/0309082].

\bibitem{Myers:1986un}
  R.~C.~Myers and M.~J.~Perry,
  Annals Phys.\  {\bf 172}, 304 (1986).

\bibitem{Kunz:2006jd}
  J.~Kunz, D.~Maison, F.~Navarro-Lerida and J.~Viebahn,
  Phys.\ Lett.\  B {\bf 639}, 95 (2006)
  [arXiv:hep-th/0606005].

\bibitem{Youm:1997hw}
  D.~Youm,
  Phys.\ Rept.\  {\bf 316}, 1 (1999)
  [arXiv:hep-th/9710046].

\bibitem{Llatas:1996gh}
  P.~M.~Llatas,
  Phys.\ Lett.\  B {\bf 397}, 63 (1997)
  [arXiv:hep-th/9605058].

\bibitem{Horowitz:1995tm}
  G.~T.~Horowitz and A.~Sen,
  Phys.\ Rev.\  D {\bf 53}, 808 (1996)
  [arXiv:hep-th/9509108].

\bibitem{Aliev:2005npa}
  A.~N.~Aliev,
  Mod.\ Phys.\ Lett.\  A {\bf 21}, 751 (2006)
  [arXiv:gr-qc/0505003].

\bibitem{Aliev:2006yk}
  A.~N.~Aliev,
  Phys.\ Rev.\  D {\bf 74}, 024011 (2006)
  [arXiv:hep-th/0604207].

\bibitem{Sheykhi:2008bs}
  A.~Sheykhi, M.~Allahverdizadeh, Y.~Bahrampour and M.~Rahnama,
  Phys.\ Lett.\  B {\bf 666}, 82 (2008)
  [arXiv:0805.4464 [hep-th]].

\bibitem{Kunduri:2004da}
  H.~K.~Kunduri and J.~Lucietti,
  Phys.\ Lett.\  B {\bf 609}, 143 (2005)
  [arXiv:hep-th/0412153].

\bibitem{NavarroLerida:2007ez}
  F.~Navarro-Lerida,
  Gen.\ Relativ.\ Gravit.\ (2010) in press 
  [arXiv:0706.0591 [hep-th]].

\bibitem{Allahverdizadeh:2010xx}
  M.~Allahverdizadeh, J.~Kunz and F.~Navarro-L\'erida,
  Phys.\ Rev.\  D {\bf 82}, 024030 (2010)
  [arXiv:1004.5050 [gr-qc]].

\bibitem{Kunz:2005nm}
  J.~Kunz, F.~Navarro-Lerida and A.~K.~Petersen,
  Phys.\ Lett.\  B {\bf 614}, 104 (2005)
  [arXiv:gr-qc/0503010].

\bibitem{Kunz:2006eh}
  J.~Kunz, F.~Navarro-Lerida and J.~Viebahn,
  Phys.\ Lett.\  B {\bf 639}, 362 (2006)
  [arXiv:hep-th/0605075].

\bibitem{Aliev:2004ec}
  A.~N.~Aliev and V.~P.~Frolov,
  Phys.\ Rev.\  D {\bf 69}, 084022 (2004)
  [arXiv:hep-th/0401095].

\bibitem{Breckenridge:1996sn}
  J.~C.~Breckenridge, D.~A.~Lowe, R.~C.~Myers, A.~W.~Peet, A.~Strominger and C.~Vafa,
  Phys.\ Lett.\  B {\bf 381}, 423 (1996)
  [arXiv:hep-th/9603078].

\bibitem{Breckenridge:1996is}
  J.~C.~Breckenridge, R.~C.~Myers, A.~W.~Peet and C.~Vafa,
  Phys.\ Lett.\  B {\bf 391}, 93 (1997)
  [arXiv:hep-th/9602065].

\bibitem{Cvetic:2004hs}
  M.~Cvetic, H.~Lu and C.~N.~Pope,
  Phys.\ Lett.\  B {\bf 598}, 273 (2004)
  [arXiv:hep-th/0406196].

\bibitem{Chong:2005hr}
  Z.~W.~Chong, M.~Cvetic, H.~Lu and C.~N.~Pope,
  Phys.\ Rev.\ Lett.\  {\bf 95}, 161301 (2005)
  [arXiv:hep-th/0506029].

\bibitem{Kunz:2005ei}
  J.~Kunz and F.~Navarro-Lerida,
  Phys.\ Rev.\ Lett.\  {\bf 96}, 081101 (2006)
  [arXiv:hep-th/0510250].

\bibitem{Kunz:2006yp}
  J.~Kunz and F.~Navarro-Lerida,
  Phys.\ Lett.\  B {\bf 643}, 55 (2006)
  [arXiv:hep-th/0610036].

\bibitem{Aliev:2008bh}
  A.~N.~Aliev and D.~K.~Ciftci,
  Phys.\ Rev.\  D {\bf 79}, 044004 (2009)
  [arXiv:0811.3948 [hep-th]].

\bibitem{Kleihaus:2002tc}
  B.~Kleihaus, J.~Kunz and F.~Navarro-Lerida,
  Phys.\ Rev.\ Lett.\  {\bf 90}, 171101 (2003)
  [arXiv:hep-th/0210197].

\bibitem{Gauntlett:1998fz}
  J.~P.~Gauntlett, R.~C.~Myers and P.~K.~Townsend,
  Class.\ Quant.\ Grav.\  {\bf 16}, 1 (1999)
  [arXiv:hep-th/9810204].





\bibitem{Gibbons:2004js}
  G.~W.~Gibbons, H.~Lu, D.~N.~Page and C.~N.~Pope,
  Phys.\ Rev.\ Lett.\  {\bf 93}, 171102 (2004)
  [arXiv:hep-th/0409155].

\bibitem{Kunz:2007jq}
  J.~Kunz, F.~Navarro-Lerida and E.~Radu,
  Phys.\ Lett.\  B {\bf 649}, 463 (2007)
  [arXiv:gr-qc/0702086].

\bibitem{Brihaye:2007vm}
  Y.~Brihaye, E.~Radu and C.~Stelea,
  Class.\ Quant.\ Grav.\  {\bf 24}, 4839 (2007)
  [arXiv:hep-th/0703046].

\bibitem{Brihaye:2008br}
  Y.~Brihaye and T.~Delsate,
  Phys.\ Rev.\  D {\bf 79}, 105013 (2009)
  [arXiv:0806.1583 [gr-qc]].

\bibitem{Aliev:2006tt}
  A.~N.~Aliev,
  Class.\ Quant.\ Grav.\  {\bf 24}, 4669 (2007)
  [arXiv:hep-th/0611205].

\bibitem{Aliev:2007qi}
  A.~N.~Aliev,
  Phys.\ Rev.\  D {\bf 75}, 084041 (2007)
  [arXiv:hep-th/0702129].

\bibitem{Sheykhi:2008rm}
  A.~Sheykhi and M.~Allahverdizadeh,
  Phys.\ Rev.\  D {\bf 78}, 064073 (2008)
  [arXiv:0809.1131 [gr-qc]].

\bibitem{Sheykhi:2009vb}
  A.~Sheykhi and M.~Allahverdizadeh,
  Gen.\ Rel.\ Grav.\  {\bf 42}, 367 (2010)
  [arXiv:0904.1776 [hep-th]].
  
\bibitem{Ghosh:2007jb}
  T.~Ghosh and S.~SenGupta,
  Phys.\ Rev.\  D {\bf 76}, 087504 (2007)
  [arXiv:0709.2754 [hep-th]].

\bibitem{Ghosh:2002ut}
  T.~Ghosh and P.~Mitra,
  Class.\ Quant.\ Grav.\  {\bf 20}, 1403 (2003)
  [arXiv:gr-qc/0212057].

\bibitem{Sheykhi:2006jh}
  A.~Sheykhi and N.~Riazi,
  Int.\ J.\ Mod.\ Phys.\  {\bf 22}, 4849 (2007)
  [arXiv:hep-th/0605042].

\bibitem{Sheykhi:2006ee}
  A.~Sheykhi and N.~Riazi,
  Int.\ J.\ Theor.\ Phys.\  {\bf 45}, 2453 (2006)
  [arXiv:hep-th/0605072].

\bibitem{Yazadjiev:2005aw}
  S.~S.~Yazadjiev,
  Phys.\ Rev.\  D {\bf 72}, 104014 (2005)
  [arXiv:hep-th/0511016].


\bibitem{Yazadjiev:2006iv}
  S.~S.~Yazadjiev,
  Phys.\ Rev.\  D {\bf 74}, 024022 (2006)
  [arXiv:hep-th/0605271].

\bibitem{Allahverdizadeh:2009ay}
  M.~Allahverdizadeh, K.~Matsuno, and A.~Sheykhi,
  Phys.\ Rev.\  D {\bf 81}, 044001 (2010)
  [arXiv:0908.2484 [hep-th]].

  

\end{thebibliography}
\end{document}